\documentstyle[12pt]{article}
\def\be{\begin{eqnarray}}
\def\ee{\end{eqnarray}}
\def\ba{\begin{array}}
\def\ea{\end{array}}
\def\p{\phi}

\def\a{\alpha}
\def\lam{\lambda}
\def\Lam{\Lambda}
\def\laI{\lambda_{1i}}
\def\laII{\lambda_{2i}}

\def\epII{\epsilon_2}
\def\epI{\epsilon_1}
\def\pa{\partial}

\def\N{{\cal N}}
\def\G{{\cal G}}
\def\B{{\cal B}}
\def\A{{\cal A}}
\def\X{{\cal X}}

\def\E{{\cal E}}

\def\D{^{(D)}}
\begin{document}
\begin{center}
{\LARGE
{Charging Interacting Rotating Black Holes\\
\vskip 0.3cm
in Heterotic String Theory
}}
\end{center}

\vskip 1.5cm

\begin{center}
{\bf \large {Alfredo Herrera--Aguilar}}
\end{center}

\vskip 1cm

\begin{center}
Theoretical Physics Department, Aristotle University of Thessaloniki\\
54006 Thessaloniki, Greece\\
and \\
Instituto de F\'\i sica y Matem\'aticas, UMSNH\\
Edificio C--3, Ciudad Universitaria, Morelia, Mich. CP 58040 M\'exico\\
e-mail: aherrera@auth.gr
\end{center}

\vskip 1cm

\begin{abstract}

We present a formulation of the stationary bosonic string sector
of the whole toroidally compactified effective field theory of the
heterotic string as a double Ernst system which, in the framework
of General Relativity describes, in particular, a pair of
interacting spinning black holes; however, in the framework of
low--energy string theory the double Ernst system can be
particularly interpreted as the rotating field configuration of
two interacting sources of black hole type coupled to dilaton and
Kalb--Ramond fields. We clarify the rotating character of the
$B_{t\varphi}$--component of the antisymmetric tensor field of
Kalb--Ramond and discuss on its possible torsion nature. We also
recall the fact that the double Ernst system possesses a discrete
symmetry which is used to relate physically different string
vacua. Therefore we apply the normalized Harrison transformation
(a charging symmetry which acts on the target space of the
low--energy heterotic string theory preserving the asymptotics of
the transformed fields and endowing them with multiple
electromagnetic charges) on a generic solution of the double Ernst
system and compute the generated field configurations for the 4D
effective field theory of the heterotic string. This
transformation generates the $U(1)^n$ vector field content of the
whole low--energy heterotic string spectrum and gives rise to a
pair of interacting rotating black holes endowed with dilaton,
Kalb--Ramond and multiple electromagnetic fields where the charge
vectors are orthogonal to each other.

\end{abstract}

\newpage
\section{Introduction}
The low energy limit of superstring theories leads to effective field
theories of massless fields that are, in fact, supergravity theories
coupled to some matter fields. One of the main reasons for constructing
classical string solutions to these theories is that
their study allows us to address standing problems in quantum gravity since
string theory is a promising candidate for a consistent quantum theory of
gravity, being free from ultra--violet divergencies that are present in
quantum gravity of point--like particles.

With regard to the gravitational field, all the uncharged black hole solutions of
general relativity describe as well (up to a good approximation) uncharged black
holes in string theory if their mass parameter is large compared to the Plank
mass, except in the vicinity of the singularity. This is owing to the fact that
the classical equation of motion for the gravitational field in the framework of
string theory consists of the Einstein's equation plus Plank correction terms.
Thus, as long as the curvature is small compared to the Plank scale, all vacuum
solutions of general relativity are approximate solutions of string theory.
However, the situation is radically different when one considers solutions of the
Einstein--Maxwell (EM) equations. In heterotic string theory, for instance, the
dilaton field couples in a non--trivial way to the electromagnetic field, so that
every electrovacuum configuration should be accompanied by a non--constant
dilaton field. It turns out that the addition of the dilaton field dramatically
changes certain properties of black holes \cite{ghs1}.

On the other side, it has been observed considerable work in
understanding the microscopic origin of the Bekenstein--Hawking
entropy of black holes using string theory approaches (for a
review see \cite{yp}). At the beginning such investigations were
mostly carried out for supersymmetric (static extremal or
near--extremal) configurations. Later on the study of string and
brane physics with no susy became an active field of research,
namely, non--BPS branes and brane--antibrane configurations have
provided considerable insight into non--perturbative aspects of
string/M--theory \cite{sen}. In the framework of this background,
it is of interest to study more complicated configurations such as
interacting black hole sets, both static and rotating, extremal
and non--extremal, neutral and charged, and coupled to other
matter fields (see, for instance, \cite{mmr}--\cite{lt} and
references therein).

The paper is organized as follows. In Sec. 2 we review the matrix
Ernst potential (MEP) formulation of the toroidally compactified
stationary low--energy heterotic string theory and recall the
normalized Harrison symmetry which is a matrix generalization of
the Harrison's B\"acklund transformation of the stationary EM
theory \cite{har}. We then present the bosonic string truncation
of this effective theory as a double Ernst system (in particular,
it represents a pair of interacting rotating black holes in the
framework of General Relativity) and point out a discrete symmetry
of this formulation that relates physically different string vacua
in Sec. 3. We further, in Sec. 4, apply the normalized Harrison
transformation (NHT) on a generic neutral double Ernst seed
solution and recover the $U(1)^n$ vector field content of the
effective field theory of the heterotic string \cite{rmf}. We also
recall the parametrization of the Ernst potentials that correspond
to Kerr black holes in the axially symmetric case. Afterwards, in
Sec. 5 we reinterpret the axisymmetric double Ernst system in the
framework of string theory as a spinning field configuration
consisting of a pair of interacting sources of black hole type
coupled to dilaton and Kalb--Ramond fields where the rotating
character of the $B_{t\varphi}$--component of the antisymmetric
tensor field is explicit. Subsequently, in Sec. 6, we perform a
NHT on this field configuration, write down the explicit
expression of the generated charged solution and study its
physical properties. We also analyze some particular cases of the
generated solution, qualitatively compare to other previously
known field configurations in certain limiting cases and discuss
on the physical properties of the considered field configurations
and on the further development of this investigation in Sec. 7.

\section{The effective action and matrix Ernst potentials}
We start with the effective action of the heterotic string at tree
level \be S\D\!=\!\int\!d\D\!x\!\mid\!
G\D\!\mid^{\frac{1}{2}}\!e^{-\p\D}\!(R\D\!+\!
\p\D_{;M}\!\p^{(D);M} \!-\!\frac{1}{12}\!H\D_{MNP}H^{(D)MNP}\!-\!
\frac{1}{4}F^{(D)I}_{MN}\!F^{(D)IMN}), \ee where \be
F^{(D)I}_{MN}\!=\!\pa_MA^{(D)I}_N\!-\!\pa _NA^{(D)I}_M, \quad\quad
H\D_{MNP}\!=\!\pa_MB\D_{NP}\!-\!\frac{1}{2}A^{(D)I}_M\,F^{(D)I}_{NP}\!+\!
\mbox{{\rm cycl. perms. of} M,\,N,\,P.}\nonumber\ee Here
$G\D_{MN}$ is the metric, $B\D_{MN}$ is the anti--symmetric
Kalb-Ramond field, $\p\D$ is the dilaton and $A^{(D)I}_M$ is a set
of $U(1)$ vector fields ($I=1,\,2,\,...,n$) and $M,N,P=1,2,...,D$.
In the consistent critical case $D=10$ and $n=16$, but we shall
leave these parameters arbitrary for the time being and will
consider the $D=4$ theory later on.  In \cite{ms} it was shown
that after the compactification of this model on a $D-3=d$--torus,
the resulting stationary theory possesses the $SO(d+1,d+1+n)$
symmetry group ($U$--duality \cite{u}) and describes effective 3D
gravity with the metric tensor \be
g_{\mu\nu}\!=\!e^{-2\p}\!\left(G\D_{\mu\nu}\!-
\!G\D_{m+3,\mu}G\D_{n+3,\nu}G^{mn}\right) \ee coupled to the
following set of three--fields:

\noindent a) scalar fields \be G\!\equiv\!G_{mn}\!=
\!G\D_{m+3,n+3},\,\,\, B\!\equiv\!B_{mn}\!= \!B\D_{m+3,n+3},\,\,\,
A\!\equiv\!A^I_m\!= \!A^{(D)I}_{m+3},\,\,\,
\p\!=\!\p\D\!-\!\frac{1}{2}{\rm ln|det}\,G|, \ee \noindent
b)antisymmetric tensor field \be
B_{\mu\nu}\!=\!B\D_{\mu\nu}\!\!-\!4B_{mn}A^m_{\mu}A^n_{\nu}\!-\!
2\!\left(A^m_{\mu}A^{m+d}_{\nu}\!-\!A^m_{\nu}A^{m+d}_{\mu}\right),
\ee \noindent c)vector fields $A^{(a)}_{\mu}=
\left((A_1)^m_{\mu},(A_2)^{m+d}_{\mu},(A_3)^{2d+I}_{\mu}\right)$
\be (A_1)^m_{\mu}\!=\!\frac{1}{2}G^{mn}G\D_{n+3,\mu},\,
(A_3)^{I+2d}_{\mu}\!=\!-\frac{1}{2}A^{(D)I}_{\mu}\!+\!A^I_nA^n_{\mu},\,
(A_2)^{m+d}_{\mu}\!=\!\frac{1}{2}B\D_{m+3,\mu}\!\!-\!B_{mn}A^n_{\mu}\!+\!
\frac{1}{2}A^I_{m}A^{I+2d}_{\mu}, \ee where the subscripts
$m,n=1,2,...,d$; $\mu,\nu=1,2,3$; and $a=1,...,2d+n$. In this
paper we set $B_{\mu\nu}=0$ to remove the effective cosmological
constant from our consideration.

All vector fields in three dimensions can be dualized on--shell
with the aid of the pseudoscalar fields $u$, $v$ and $s$ as
follows:
\begin{eqnarray}
\nabla\times\overrightarrow{A_1}&=&\frac{1}{2}e^{2\p}G^{-1}
\left(\nabla u+(B+\frac{1}{2}AA^T)\nabla v+A\nabla s\right),
\nonumber                          \\
\nabla\times\overrightarrow{A_3}&=&\frac{1}{2}e^{2\p} (\nabla
s+A^T\nabla v)+A^T\nabla\times\overrightarrow{A_1},
\label{dual}\\
\nabla\times\overrightarrow{A_2}&=&\frac{1}{2}e^{2\p}G\nabla v-
(B+\frac{1}{2}AA^T)\nabla\times\overrightarrow{A_1}+
A\nabla\times\overrightarrow{A_3}. \nonumber
\end{eqnarray}
Thus, the effective stationary theory describes gravity
$g_{\mu\nu}$ coupled to the scalars $G$, $B$, $A$, $\p$ and
pseudoscalars $u$, $v$, $s$. These matter fields can be arranged
in the following pair of matrix Ernst potentials: \be \X=\left(
\ba{cc} -e^{-2\p}+v^TXv+v^TAs+\frac{1}{2}s^Ts&v^TX-u^T \cr
Xv+u+As&X \ea \right), \quad \qquad\A=\left( \ba{c} s^T+v^TA \cr A
\ea \right), \label{XA}\ee where $X=G+B+\frac{1}{2}AA^T$. These
matrices have dimensions $(d+1) \times (d+1)$ and $(d+1) \times
n$, respectively, and their components have the following physical
meaning: the relevant information about the gravitational field is
contained in the matrix potential $X$ through the matrix $G$,
whereas its rotational character is encoded in the dualized
variable $u$; $X$ also parameterizes the antisymmetric
Kalb--Ramond tensor field $B$, whereas its multidimensional
components are dualized through $v$; the 3D dilaton is $\phi$, and
$A$ and $s$ stand for electric and magnetic potentials,
respectively. In terms of the MEP the effective stationary theory
adopts the form \be ^3S\!= \!\int\!d^3x\!\mid
g\mid^{\frac{1}{2}}\!\{\!-\!R\!+ \!{\rm
Tr}[\frac{1}{4}\left(\nabla \X\!-\!\nabla \A\A^T\right)\!\G^{-1}
\!\left(\nabla \X^T\!-\!\A\nabla \A^T\right)\!\G^{-1}
\!+\!\frac{1}{2}\nabla \A^T\G^{-1}\nabla \A]\},\label{acXA}\ee
where $\X=\G+\B+\frac{1}{2}\A\A^T$, then \,
$\G=\frac{1}{2}\left(\X+\X^T-\A\A^T\right)$ and \be \G= \left(
\ba{cc} -e^{-2\p}+v^TGv&v^TG \cr Gv&G \ea \right), \quad \B=\left(
\ba{cc} 0&v^TB-u^T \cr Bv+u&B \ea \right). \ee

In \cite{hk3} it was shown that there exist a map between the
stationary actions of the low--energy heterotic string and
Einstein--Maxwell theories. The map reads \be
\X\longleftrightarrow -E, \quad \A\longleftrightarrow F, \nonumber
\ee and also involves an interchange of the {\it matrix
transposition} operation with the {\it complex conjugation} one.
Here $E$ and $F$ are the complex Ernst potentials (gravitational
and electromagnetic, respectively) of the stationary EM theory
\cite {e}. This map allows us to extrapolate the results obtained
in the EM theory to the heterotic string one using the MEP
formulation.
\subsection{The normalized Harrison transformation}
In \cite{hk3} it was shown that the total symmetry group
$SO(d+1,d+1+n)/[SO(d+1)\times SO(d+n+1)]$ of the stationary
low--energy effective field theory of the heterotic string can be
decomposed into scaling, electromagnetic rotation, gauge and
non--linear (Ehlers and Harrison) matrix transformations in terms
of the MEP, just as this takes place in the stationary EM theory.
Further, following the work of Kinnersley \cite{kin} for the
stationary EM theory, the symmetries of the stationary action
(\ref{acXA}) were classified according to their charging
properties in \cite{hk5}. It turns out that only the generalized
Ehlers and Harrison transformations act in a non--trivial way on a
given seed space--time since they introduce physical charges into
the generated solution (by applying a scaling transformation on a
seed field configuration we just make it larger or smaller without
adding any charge and, in this sense, without changing its
physical properties, the electromagnetic rotation leaves unchanged
a charged configuration; finally, the gauge symmetries of the MEP
do not alter a given seed solution).

At this point we would like to mention that the NHT constitutes a
matrix generalization of the charging symmetry of B\"acklund type
introduced by Harrison in the framework of the EM theory
\cite{har}. In that context, by applying the Harrison
transformation on a neutral seed solution one introduces a complex
parameter which endows with electric and/or magnetic charges the
generated field configuration. On the other side, by applying the
normalized Ehlers transformation on a given seed solution, one
does not introduce charges of electromagnetic nature since
starting with a neutral configuration one ends again with a
neutral solution (within the context of General Relativity, it
introduces the NUT parameter).

Thus, in the framework of the effective field theory of the
heterotic string, under the NHT we can construct charged string
vacua from neutral ones with the same asymptotical values of the
fields, just as this takes place in the framework of General
Relativity where we obtain the Kerr--Newman black hole from the
Kerr one under such non--linear symmetry. However, in the
low--energy string theory context, the NHT introduces $(d+1)\times
n$ real parameters through a matrix $\lambda$ and generates a
field configuration endowed with $(d+1)\times n$ electromagnetic
charges starting from a neutral seed solution. In the concrete
case of the critical effective field theory of the heterotic
string, when $D=10$ and $n=16$, it generates $128$ electromagnetic
charges.

In this respect, the non--linear NHT selects from the total
$SO(d+1,d+1+n)/[SO(d+1)\times SO(d+n+1)]$ symmetry group of the
stationary effective field theory of the heterotic string, the
symmetries that endow a given seed field configuration with
electromagnetic charges. Another important issue of this method
concerns its applicability to stationary systems only, thus, it
cannot be implemented to effective field theories depending on
more than three space--time dimensions.

In this respect we would like to point out that the solution
generating technique implemented in the present work is quite
different from the method based in orthogonal transformations that
has been applied in \cite{ms} and \cite{hassen}, for instance.
However, it seems that the result of applying both techniques is
quite similar and in some cases, the same. It is interesting to
study the relationship between both solution generating methods.

Thus, the matrix transformation \be
&&\A\rightarrow\left(1+\frac{1}{2}\Sigma\lambda\lambda^T\right)
\left(1-\A_0\lambda^T+\frac{1}{2}\X_0\lambda\lambda^T\right)^{-1}
\left(A_0-\X_0\lambda\right)+\Sigma\lambda,
\\
&&\X\rightarrow\left(1+\frac{1}{2}\Sigma\lambda\lambda^T\right)
\left(1-\A_0\lambda^T+\frac{1}{2}\X_0\lambda\lambda^T\right)^{-1}
\left[\X_0+\left(\A_0-\frac{1}{2}\X_0\lambda\right)\lambda^T\Sigma\right]
+\frac{1}{2}\Sigma\lambda\lambda^T\Sigma, \nonumber \ee where
$\Sigma=diag(-1,-1;1,1,...,1)$ and $\lambda$ is an arbitrary
constant $(d+1)\times n$--matrix parameter, generates charged
string solutions (with non--zero potential $\A$) from neutral ones
if we start from the seed potentials \be \X_0\ne 0 \qquad
\mbox{\rm and} \qquad \A_0=0.\nonumber\ee This solution generation
procedure allows us to generate the $U(1)^n$ electromagnetic
spectrum of the effective field theory of the heterotic string
starting with just field spectrum of the low--energy bosonic
string theory.

\section{Bosonic string truncation vs double Ernst system}
Let us consider just the low--energy bosonic string sector of the
whole effective stationary heterotic string  theory. Thus, we must
set to zero all the $U(1)$ field strengths which correspond to the
winding modes of the reduced theory; in terms of the MEP this is
equivalent to dropping the matrix $\A$ in the action (\ref{acXA})
\be ^3S\!= \!\int\!d^3x\!\mid
g\mid^{\frac{1}{2}}\!\left\{\!-\!R\!+ \frac{1}{4}\!{\rm
Tr}\,\left[ \nabla \X\!\G^{-1}\!\nabla
\X^T\!\G^{-1}\right]\right\} =\!\int\!d^3x\!\mid
g\mid^{\frac{1}{2}}\!\left\{\!-\!R\!+ \frac{1}{4}\!{\rm
Tr}\,\left(J^{\X}J^{\X^{T}}\right)\right\} \label{acX}\ee where
now $\X=\G+\B$,\,\, $\G=\frac{1}{2}\left(\X+\X^T\right)$ and
$J^{\X}=\nabla \X\G^{-1}$.

There are several physically different theories (and truncations)
that can be expressed by the action (\ref{acX}), and hence admit a
double Ernst formulation (see, for instance,
\cite{gk2}--\cite{hkdos}); among them we find the $D=4$
low--energy bosonic string theory, the bosonic sector of $D=4$,
${\cal N}=4$ supergravity, the $D=5$ Einstein--Kalb--Ramond
effective model, where the dilaton field is set to zero, etc. In
these cases $\G$ and $\B$ are $2\times 2$--matrices and can be
parameterized in the following form \be \G= \frac{p_1}{p_2} \left(
\ba{cc} 1&q_2\cr q_2&p_2^2+q_2^2 \ea \right), \qquad \qquad \B=
q_1 \left(
\begin{array}{crc}
0  & \quad & -1\\
1 & \quad & 0\\
\end{array}
\right). \label{GB} \ee Thus, the action of the matter fields
takes the form
\begin{eqnarray}
^3 S_m = \frac{1}{2} \int d^3x {\mid g \mid}^{\frac {1}{2}}
\left\{ p_1^{-2}\left[(\nabla p_1)^2 + (\nabla q_1)^2\right] +
p_2^{-2}\left[(\nabla p_2)^2 + (\nabla q_2)^2\right] \right\},
\label{pq}
\end{eqnarray}
which allows us to introduce two independent to each other Ernst--like potentials
\begin{equation}
\epsilon _1 =p_1+iq_1, \qquad \epsilon _2=p_2+iq_2.
\end{equation}
In terms of these field variables, the action of the system can be rewritten as
a double Ernst system in the K\"ahler form \cite{mazur}:
\begin{eqnarray}
^3 S =
\int d^3x {\mid g \mid}^{\frac {1}{2}} \left\{ - ^3 R +
2\left(J^{\epsilon _1}J^{\overline\epsilon _1} +
J^{\epsilon _2}J^{\overline\epsilon _2} \right) \right\},
\label{acernst}
\end{eqnarray}
where $J^{\epsilon _k}=\nabla\epsilon _k\,(\epsilon
_k+\overline\epsilon _k)^{-1}$ and $k=1,2$. In the framework of
the stationary General Relativity each complex Ernst potential
describes, in general, a spinning field configuration and, in
particular, an axially symmetric rotating black hole (or
spherically symmetric static black hole when the potential is
real). Thus, the action (\ref{acernst}) represents the linear
superposition of two rotating black holes among other
configurations.

A mathematically equivalent, but physically different $2 \times
2$--matrix representation arises from (\ref{pq}) by making use of
the discrete symmetry $p_1 \longleftrightarrow p_2$, \,
$q_1\longleftrightarrow q_2$. This fact allows us to define the
new matrices $\G'$ and $\B'$, and hence the new matrix potential
$\X'=\G'+\B'$, and to write down the action that corresponds to
the primed quantities: \be ^3S\!= \!\int\!d^3x\!\mid
g\mid^{\frac{1}{2}}\!\left\{\!-\!R\!+ \frac{1}{4}\!{\rm
Tr}\,\left(J^{\X'}J^{\X'^{T}}\right)\right\} \!=\!\int\!d^3x\!\mid
g\mid^{\frac{1}{2}}\!\left\{\!-\!R\!+ 2\left(J^{\epsilon
'_1}J^{\overline\epsilon '_1}+ J^{\epsilon
'_2}J^{\overline\epsilon '_2}\right)\right\},\label{acx'} \ee
where similarly $J^{\X'}=\nabla\X'\G'^{-1}$, \, $J^{\epsilon
'_k}=\nabla\epsilon '_k\,(\epsilon '_k+\overline\epsilon
'_k)^{-1}$,\, $\epsilon'_1=p_2+iq_2$ and $\epsilon '_2=p_1+iq_1$.

In terms of the MEP the above--mentioned discrete transformation
reads $\X \longleftrightarrow \X'$; thus, the matrices $\G'$ and
$\B'$ must be interpreted as new Kaluza--Klein and Kalb--Ramond
fields. This symmetry mixes the gravitational and matter degrees
of freedom of the theories. It works like the Bonnor
transformation of the Einstein--Maxwell theory \cite{b}, but in
the bosonic string realm. It can be used to generate new solutions
starting from seed solutions as pure Kaluza--Klein or
Kaluza--Klein--Dilaton string vacua.
\section{Charging double Ernst solutions in string theory}
For concreteness, in this Sec. we shall consider the truncated
$D=4$ action of the effective field theory of the bosonic string
with a non--null timelike Killing vector; thus, the vectors fields
of the effective field theory of the heterotic string encoded in
the MEP $\A$ vanish and the stationary low--energy action is
effectively described by (\ref{acX}) with $\G$ and $\B$ defined by
(\ref{GB}). Subsequently, we will generate the electromagnetic
sector ($U(1)^n$ vector field content) of the low--energy
heterotic string theory via the NHT and, finally, will impose
axial symmetry in order to consider the MEP corresponding to a
couple of interacting Kerr black holes.

The seed MEP that correspond to the neutral stationary double
Ernst system are \be \X_0= \left( \ba{ccc}
\frac{p_1}{p_2}&\quad&\frac{p_1q_2-q_1p_2}{p_2}\cr
\quad&\quad&\quad\cr \frac{p_1q_2+q_1p_2}{p_2}&\quad&
\frac{p_1}{p_2}(p_2^2+q_2^2) \ea \right), \qquad \qquad \A_0=0.
\label{seedpots}\ee For the simplest case the charge matrix $\lam$
that parameterizes the NHT is conformed by two vector rows as
follows \be \lam= \left( \ba{cccc} \lam_{11}&\lam_{12}& ...
&\lam_{1n}\cr \lam_{21}&\lam_{22}& ... &\lam_{2n}\cr \ea
\right),\label{lambda2}\ee where $n\ge 2$ for consistency; the
parameters $\laI$ and $\laII$ can be interpreted as the
electromagnetic charges of the generated field configuration. When
$n=6$ the generated field spectrum corresponds to the bosonic
sector of ${\cal N}=4$, $D=4$ supergravity; however, here we shall
leave it arbitrary for the sake of generality. After applying the
NHT on the generic double Ernst seed solution (\ref{seedpots})
with the matrix $\lam$ (\ref{lambda2}), the transformed MEP read
\be \X_{11}\!=\!\frac{1}{\Xi}\left[\left(4+\Lam^2 |\epII
|^2\right) Re\epI+2\left(\laI^2+\laII^2 |\epI |^2\right)Re\epII -
4\laI\laII Re\epII Im\epI\right], \ee \be
\X_{12}\!=\!\frac{1}{\Xi}\left\{\Gamma_{-}\left( Re\epI
Im\epII-Re\epII Im\epI\right)+2\laI\laII\left[ (1-|\epII
|^2)Re\epI +(1-|\epI |^2)Re\epII\right]\right\}, \ee \be
\X_{21}\!=\!\frac{1}{\Xi}\left\{\Gamma_{+} \left(Re\epI
Im\epII+Re\epII Im\epI\right)+2\laI\laII\left[ (1-|\epI
|^2)Re\epII -(1-|\epII |^2)Re\epI\right]\right\}, \ee \be
\X_{22}\!=\!\frac{1}{\Xi}\left[ \left(\Lam^2 +4|\epII |^2\right)
Re\epI+2\left(\laII^2+\laI^2 |\epI |^2\right)Re\epII + 4\laI\laII
Re\epII Im\epI\right], \ee \be \A_{1j}\!=\!\frac{-2}{\Xi}\left\{
\left[\left(2+\laII^2 |\epII |^2\right)Re\epI + \left(2+\laII^2
|\epI |^2\right)Re\epII + \laI\laII\left(Re\epI Im\epII-Re\epII
Im\epI\right)\right]\lam_{1j}+\right. \nonumber \ee \be
\left.\left[\left(2-\laI^2\right)\left(Re\epI Im\epII-Re\epII
Im\epI\right)- \laI\laII\left(|\epII |^2Re\epI +|\epI
|^2Re\epII\right)\right]\lam_{2j} \right\}, \ee \be
\A_{2j}\!=\!\frac{-2}{\Xi}\left\{\left[\left(2-\laII^2\right)
\left(Re\epI Im\epII+Re\epII Im\epI\right)-\laI\laII \left(Re\epI
+|\epI |^2 Re\epII\right)\right]\lam_{1j}\right.+ \nonumber \ee
\be \left.\left[\left(\laI^2+2|\epII |^2\right)Re\epI +
\left(2+\laI^2 |\epI |^2\right)Re\epII + \laI\laII\left(Re\epI
Im\epII+Re\epII Im\epI\right)\right]\lam_{2j}\right\}, \ee where
$\Xi=2\left(\laI^2+\laII^2 |\epII |^2\right)Re\epI +
\left(4+\Lam^2 |\epI |^2\right)Re\epII + 4\laI\laII Re\epI
Im\epII$, $\Gamma_{-}=4-2\laI^2+2\laII^2-\Lam^2$,
$\Gamma_{+}=4+2\laI^2-2\laII^2-\Lam^2$ and
$\Lam^2=\laI^2\lambda_{2j}^2-\left(\laI\laII\right)^2$. It is
straightforward to check that the generated MEP preserve the
asymptotic trivial values of the seed MEP $\X_0$ and $\A_0$.

It remains just to consider concrete seed solutions to the double
Ernst system, to substitute the expressions for $\epI$ and $\epII$
in the transformed MEP formulae, to compute the original fields of
the theory and to give an interpretation of the generated families
of solutions.
\subsection{Axisymmetric case and interacting Kerr black holes}
Since most of the physically meaningful solutions of General
Relativity that can be used as seed solutions in our approach are
axisymmetric, we shall impose axial symmetry and write the line
element in the Lewis--Papapetrou form using canonical Weyl
coordinates \be ds^2=
\left(dx^m+2(A_1)^m_{\varphi}d\varphi\right)^T
G_{mn}\left(dx^n+2(A_1)^n_{\varphi}d\varphi\right)+
e^{2\p}\left[e^{2\gamma}\left(d\rho^2+dz^2\right)+\rho^2d\varphi^2\right],
\ee where $\gamma$, $\p$, $G_{mn}$ and $(A_1)^m_{\varphi}$ are
$\varphi$--independent. A solution of our axisymmetric system can
be constructed using the solutions of the double vacuum Einstein
equations written in the Ernst form in terms of $\epsilon _k$ and
$\gamma ^{\epsilon _k}$
\be \nabla (\rho J^{\epsilon _k})&=& \rho
J^{\epsilon _k}(J^{\epsilon _k}-J^{\bar \epsilon _k}),
\nonumber\\
\partial _{z} \gamma ^{\epsilon _k}&=&
\rho \left [(J^{\epsilon _k})_z (J^{\bar \epsilon _k})_{\rho}
+(J^{\bar \epsilon _k})_z (J^{\epsilon _k})_{\rho}\right ],\\
\partial _{\rho} \gamma ^{\epsilon _k}&=&
\rho \left [{|(J^{\epsilon _k})_{\rho}}|^2
-{|(J^{\epsilon _k})_z}|^2\right],
\nonumber
\ee
if one identifies the function $\gamma$, that accounts for the general
relativistic interaction between de black holes, in the following way
$\gamma \equiv \gamma ^{\epsilon _1} + \gamma ^{\epsilon _2}$ because
of the linear character that possesses the superposition of the sources
in the action (\ref{acernst}).

Let us consider the double Kerr solution as the seed one.
The Ernst potentials corresponding to two Kerr black holes with
sources in different points of the symmetry axis are:
\begin{eqnarray}
\epsilon_k = 1 - \frac {2m_k}{r_k + ia_k\cos\theta_k},
\label{kerrpots}
\end{eqnarray}
where $m_k$ and $a_k$ are constant parameters which define the masses and
rotations of the sources of the Kerr field configurations.
Weyl and Boyer--Lindquist coordinates are related through
\begin{eqnarray}
\rho=\sqrt{(r_k - m_k)^2-\sigma _k^2}\sin\theta_k,
\qquad
z = z_k + (r_k - m_k)\cos\theta_k,
\end{eqnarray}
where the sources are located at $z_k$ and $\sigma _k^2=m_k^2-a_k^2$.
Thus, for the function $\gamma^{\epsilon_k}$ we have
$e^{2\gamma^{\epsilon_{_k}}}=P_k/Q_k$,
where $P_k = \Delta_k - a_k^2 \sin^2 \theta _k$, \, $Q_k = \Delta_k +
\sigma _k^2 \sin^2 \theta _k$ and $\Delta _k = r_k^2 - 2m_k r_k + a_k^2$.
\section{Seed solutions of $D=4$ heterotic string theory}
Let us consider some particular solutions for the stationary
action of the $D=4$ effective field theory of the bosonic string.
In this case the three--dimensional field configurations of the
low--energy theory are parameterized in terms of the double Ernst
system as follows \be
G_{tt}=\frac{Re\epsilon_1}{Re\epsilon_2}\mid\epsilon_2\mid^2,
\qquad B\equiv0, \qquad e^{-2\p}=-\frac{Re\epsilon_1
Re\epsilon_2}{\mid \epsilon_2\mid ^2}, \qquad u=Im\epsilon_1,
\qquad v=\frac{Im\epsilon_2}{\mid\epsilon_2\mid^2},
\label{seedsol}\ee In particular, if we consider the case when the
Ernst potentials correspond to a double Kerr black hole, the 4D
field configuration in the string frame reads \be
ds_4^2=G_{tt}\left(dt+\omega_{\varphi}d\varphi\right)^2+
\frac{e^{2\p^{(4)}}}{\mid G_{tt}\mid}
\left[e^{2\gamma}\left(d\rho^2+dz^2\right)+\rho^2d\varphi^2\right],
\label{metric} \ee \be G_{tt}=-\frac{P_1R_2}{\tilde
r_1P_2},\qquad\quad \omega_{\varphi}\equiv 2(A_1)_{\varphi}
=\frac{2m_1\a_1r_1\sin^2\theta_1}{P_1},\qquad\quad
e^{2\gamma}=\frac{P_1P_2}{Q_1Q_2},\nonumber\ee \be
e^{\p^{(4)}}=\frac{R_2}{P_2},\qquad\quad B_{t\varphi}\equiv
2(A_2)_{\varphi} =\frac{2m_2\a_2\sin^2\theta_2(r_2-2m_2)}{P_2},
\label{matter} \ee where $R_k=(r_k-2m_k)^2+\a_k^2\cos^2\theta_k$
and $\tilde r_k=r_k^2+\a_k^2\cos^2\theta_k$, and describe the
axisymmetric rotating field configuration of two interacting
sources $m_1$ and $m_2$ of black hole type located at $z_1$ and
$z_2$ respectively, and endowed with dilaton and Kalb--Ramond
fields. It is worth noting that the $G_{tt}$--component of the
metric tensor defines the horizon of each gravitational source,
however, the quantity $\omega_{\varphi}$ parameterizes only the
rotation of the source $m_1$ since it does not depend on the
parameter $\a_2$. Interestingly, the Kalb--Ramond field possesses
a rotational character (compare it to $\omega_{\varphi}$), depends
just on the parameter $\a_2$ and has no Coulomb term in its
asymptotical decomposition. Thus, the rotating character of the
gravitational field generated by the source $m_2$, in the
framework of General Relativity, effectively translates into a
`rotating' antisymmetric background in the framework of
low--energy string theory. This fact could be related to the {\it
torsion} nature of the $B_{t\varphi}$--component of the
antisymmetric tensor field of Kalb--Ramond, however, these
interpretation deserves a separate investigation and will be
pursued elsewhere. Finally, let us point out that the 4D dilaton
field is located at $z_2$. The explicit expressions for
$(A_1)_{\varphi}$ and $(A_2)_{\varphi}$ have been computed using
the dualization formulae (\ref{dual}) with the aid of the
following coordinate transformation
\begin{eqnarray}
2r_i&=&\sqrt{\rho^2+\left(z-z_i+\sqrt{m_i^2-\alpha_i^2}\right)^2}+
\sqrt{\rho^2+\left(z-z_i-\sqrt{m_i^2-\alpha_i^2}\right)^2}+2m_i,\\
2\cos\theta_i&=&\left(\sqrt{\rho^2+\left(z-z_i+\sqrt{m_i^2-\alpha_i^2}\right)^2}-
\sqrt{\rho^2+\left(z-z_i-\sqrt{m_i^2-\alpha_i^2}\right)^2}\right)/\sqrt{m_i^2-\alpha_i^2}.
\label{rtetavsrozeta} \nonumber\end{eqnarray}

Since at spatial infinity $G_{tt}\mid_{\infty}\longrightarrow -1$,
either $\epsilon_1$ or $\epsilon_2$ must adopt the asymptotic
value $-1$; for concreteness we have chosen
$\epsilon_1\mid_{\infty}\longrightarrow -1$. Looking at the
asymptotical behaviour of the $G_{tt}$ component of the
gravitational tensor we get \be G_{tt}\mid_{\infty}\sim
-1+\frac{2(m_1+m_2)}{\rho}, \nonumber \ee as expected since the
double black hole is generated by the masses $m_1$ and $m_2$; on
the other side, for the dilaton field we asymptotically have the
following relation \be \varphi\mid_{\infty}\sim
\frac{(m_1-m_2)}{\rho} \equiv\frac{d}{\rho}, \nonumber \ee which
defines its charge. We see that this three--dimensional dilaton
charge effectively vanishes in the case when $m_1=m_2=m$.

\noindent Due to the discrete symmetry
$\epsilon_1\!\!\leftrightarrow\!\!\epsilon_2$, there is an
alternative parameterization of the seed fields \be G'_{tt}=
\frac{Re\epsilon_2}{Re\epsilon_1}\mid\epsilon_1\mid^2, \qquad
B'\equiv0, \qquad e^{-2\p'}=\frac{-Re\epsilon_1 Re\epsilon_2}{\mid
\epsilon_1\mid ^2}, \qquad v'=
\frac{Im\epsilon_1}{\mid\epsilon_1\mid^2}, \qquad u'=
Im\epsilon_2, \ee where we observe certain duality between the
antisymmetric Kalb--Ramond field and the non--diagonal term
$G_{t\varphi}$, responsible for the rotation of the gravitational
field. This relationship was already pointed out in \cite{hkdos}
and \cite{hk9} and can be seen by switching off one of the
potentials, say $\epII =1$; thus, the resulting configuration
corresponds to a single 4D rotating black hole with vanishing
dilaton and Kalb--Ramond fields. Under the discrete symmetry
mentioned above, this solution is mapped into a static field
configuration endowed with non--trivial dilaton and Kalb--Ramond
fields. It it interesting to study in more detail the string vacua
that are related via this discrete duality and establish its
physical implications (for explicit examples see \cite{hncqg}).
\section{Charged field configurations in heterotic string theory}
After applying the NHT on the seed solution (\ref{seedsol}) we get
the following three--dimensional fields: \be G\equiv
G_{tt}=\X_{22}-\frac{1}{2}\A_{2j}^2, \qquad B\equiv0, \qquad
A=\A_{2j}, \qquad
v=\frac{1}{2G}\left(\X_{12}+\X_{21}-\A_{1j}\A_{2j}\right),
\nonumber \ee \be u=v\X_{22}-\X_{12},  \qquad
s^T=\A_{1j}-v\A_{2j}, \qquad
e^{-2\p}=\frac{1}{2}\left[v(\X_{12}+\X_{21})+\A_{ij}s\right]-\X_{11}.
\ee where the appearance of the electromagnetic potential is
obvious. Similar relations hold for the primed field system that
arises under the interchange $\epI\longleftrightarrow \epII$.

With the aid of the dualization relations (\ref{dual}) we get
explicit expressions for the non--trivial components of the vector
fields \be \omega_{\varphi}\equiv 2(A_1)_{\varphi}
=\left.\left[-4\chi_{11}+2\lam_{1i}^2\chi_{21}+\left(
\lam_{1i}^2\lam_{2j}^2-(\lam_{1i}\lam_{2i})^2\right)\chi_{12}-
2\lam_{2i}^2\chi_{22}+4\lam_{1i}\lam_{2i}\chi_{23}\right]\right/DQ,
\nonumber \ee \be B_{t\varphi}\equiv 2(A_2)_{\varphi}
=\left.\left[-2\lam_{2i}^2\chi_{11}
+\left(\lam_{1i}^2\lam_{2j}^2-(\lam_{1i}\lam_{2i})^2\right)\chi_{21}
+2\lam_{1i}^2\chi_{12}
-4\chi_{22}-4\lam_{1i}\lam_{2i}\chi_{13}\right]\right/DQ,
\nonumber \ee \be (A_3)^{I}_{\varphi}
=(DQ)^{-2}\left\{\lam_{1i}\lam_{2i}\left[DQ\left(\chi_{12}+\chi_{21}\right)-
\left(4-2\lam_{2i}^2\right)\chi_{22}+\left(\lam_{1i}^2\lam_{2j}^2-
2\lam_{1i}^2-(\lam_{1i}\lam_{2i})^2\right)\chi_{21}-\right.\right.
\nonumber \ee \be
\left.\left.4\,\lam_{1i}\,\lam_{2i}\,\chi_{13}\right]+
2\left[\left(1-\lam_{1i}^2\right)\left(2-\lam_{2j}^2\right)^2-
\left(1-\lam_{2i}^2\right)(\lam_{1j}\,\lam_{2j})^2\right]
\left(\chi_{13}-\chi_{23}\right)\right\}\lam_{1I}+ \nonumber \ee
\be (DQ)^{-2}\left\{2DQ\left(\chi_{11}+\chi_{22}\right)-
DQ\lam_{1i}^2\left(\chi_{12}+2\chi_{21}\right)+
2DQ\lam_{1i}\lam_{2i}\left(\chi_{13}-\chi_{23}\right)-\right.
\nonumber \ee \be \left.
2\lam_{1i}\lam_{2i}\left[\lam_{1j}\lam_{2j}\left(\chi_{21}+\chi_{22}\right)+
\left(4-\lam_{1j}^2-\lam_{2j}^2\right)\chi_{13}+
\left(\lam_{2j}^2-\lam_{1j}^2\right)\chi_{23}\right]\right\}\lam_{2I},
\ee where $I=1,2,...n$ and the functions $\chi_{kl}$ ($l=1,2,3;$)
read \be \chi_{k1}= \frac{\a_km_kr_k\sin^2\theta_k}{P_k},\quad
\chi_{k2}= \frac{\a_km_k(r_k-2m_k)\sin^2\theta_k}{P_k},\quad
\chi_{k3}= \frac{m_k(r_k^2-2m_kr_k+\a_k^2)\cos\theta_k}{P_k},
\nonumber \ee and $DQ=4-2\lam_{1i}^2-2\lam_{2i}^2+
\lam_{1i}^2\lam_{2j}^2-(\lam_{1i}\lam_{2i})^2$. It is worth
noticing that even though the generated MEP are asymptotically
flat, the $\chi_{k3}$ functions do not vanish at spatial infinity
(they involve the so--called NUT parameters of the gravitational
field and the Dirac strings of the magnetic components of the
vector fields). Thus, in order to get asymptotically flat
gravitational field configurations, which include charged black
holes, we must get rid of the terms proportional to $\chi_{k3}$.
This can be done by imposing the orthogonality condition on the
pair of charge vectors $\lam_{1i}$ and $\lam_{2i}$: \be
\lam_{1i}\lam_{2i}=0, \label{ort} \ee however, even with this
restriction the magnetic components of the gauge fields
$(A_3)^I_{\varphi}$ still present the Dirac string singularity; in
order to remove it we need to normalize the charge vector
$\lam_{1i}$ as follows $\lam_{1i}^2=1$, but we shall leave it
unnormalized since these singularities could correspond to Dirac
strings of gravitating monopole solutions. Thus, after imposing
(\ref{ort}), the transformed metric preserves its form
(\ref{metric}) with the following fields in the string frame \be
G_{tt}=-\frac{(2-\lam_{2i}^2)^2P_1P_2 \left[4R_2\tilde
r_1+\lam_{1j}^4R_1\tilde r_2-4\lam_{1j}^2\left(P_1P_2-
4m_1m_2\a_1\a_2\cos\theta_1\cos\theta_2\right)\right]}
{\left[P_2\left(4\tilde r_1+\lam_{1j}^2\lam_{2i}^2R_1\right)
-2P_1\left(\lam_{1j}^2\tilde r_2+\lam_{2i}^2R_2\right)\right]^2},
\nonumber \ee \be e^{\p^{(4)}}
=\frac{\left(2-\lam_{2i}^2\right)\left[4R_2\tilde r_1
+\lam_{1j}^4R_1\tilde r_2-4\lam_{1j}^2\left(P_1P_2-
4m_1m_2\a_1\a_2\cos\theta_1\cos\theta_2\right)\right]}
{\left(2-\lam_{1j}^2\right)\left[-2P_1\left(\lam_{1j}^2 \tilde
r_2+\lam_{2i}^2R_2\right)+P_2\left(4\tilde r_1+
\lam_{1j}^2\lam_{2i}^2R_1\right)\right]}, \nonumber \ee \be
\omega_{\varphi}\equiv 2(A_1)_{\varphi}
=2\left.\left(4\chi_{11}-2\lam_{1j}^2\chi_{21}-
\lam_{1j}^2\lam_{2i}^2\chi_{12}+2\lam_{2i}^2\chi_{22}\right)\right/
\left[\left(2-\lam_{1j}^2\right)\left(2-\lam_{2i}^2\right)\right],
\nonumber \ee \be B^{(4)}_{t\varphi}\equiv 2(A_2)_{\varphi}
=2\left.\left(2\lam_{2i}^2\chi_{11}-\lam_{1j}^2\lam_{2i}^2\chi_{21}
-2\lam_{1j}^2\chi_{12}+4\chi_{22}\right)\right/
\left[\left(2-\lam_{1j}^2\right)\left(2-\lam_{2i}^2\right)\right],
\nonumber \ee \be
A^I=\frac{4(2-\lam_{2i}^2)^2\left(m_2\a_2\cos\theta_2P_1+
m_1\a_1\cos\theta_1P_2\right)\lam_{1I}} {\left[P_2\left(4\tilde
r_1+\lam_{1j}^2\lam_{2i}^2R_1\right)- 2P_1\left(\lam_{1j}^2\tilde
r_2+\lam_{2i}^2R_2\right)\right]}+ \nonumber \ee \be
\frac{2\left[P_1\left(\lam_{1j}^2\tilde r_2+2R_2\right)-
P_2\left(2\tilde r_1+\lam_{1j}^2R_1\right)\right]\lam_{2I}}
{\left[P_2\left(4\tilde r_1+\lam_{1j}^2\lam_{2i}^2R_1\right)-
2P_1\left(\lam_{1j}^2\tilde r_2+\lam_{2i}^2R_2\right)\right]},
\nonumber \ee \be (A_3)^I_{\varphi}
=\left.4\left(1-\lam_{1j}^2\right)\left(\chi_{23}-\chi_{13}\right)
\lam_{1I}\right/\left(2-\lam_{1j}^2\right)^2- \nonumber \ee \be
2\left.\left[2\left(\chi_{11}+\chi_{22}\right)-\lam_{1j}^2
\left(\chi_{12}+2\chi_{21}\right)\right]\lam_{2I}\right/
\left[\left(2-\lam_{1j}^2\right)\left(2-\lam_{2i}^2\right)\right],
\ee where $e^{2\gamma}$ remains the same. It is straightforward to
check that when one sets to zero the charge vectors $\lambda_{ki}$
one recovers the seed field configuration
(\ref{metric})--(\ref{matter}).

By switching to the Einstein frame, where most of the calculations
for gravitational and thermodynamical analysis are performed, we
obtain \be
ds_E^2=e^{-\p^{(4)}}ds_{st}^2=-f\left(dt-\omega_{\varphi}\right)^2+f^{-1}
\left[e^{2\gamma}\left(d\rho^2+dz^2\right)+\rho^2d\varphi^2\right],
\ee where the component of the metric tensor $f$ reads \be
f=\frac{\left(2-\lam_{1j}^2\right)\left(2-\lam_{2i}^2\right)P_1P_2}
{\left[P_2\left(4\tilde r_1+\lam_{1j}^2\lam_{2i}^2R_1\right)
-2P_1\left(\lam_{1j}^2\tilde r_2+\lam_{2i}^2R_2\right)\right]},
\nonumber \ee and has the following asymptotic behaviour \be
f\mid_{\infty}\sim 1-\frac{2\left(M_1+M_2\right)}{r}, \quad
\mbox{\rm with} \quad
M_1\equiv\frac{\left(4-\lam_{1j}^2\lam_{2i}^2\right)m_1}
{\left(2-\lam_{1j}^2\right)\left(2-\lam_{2i}^2\right)}, \quad
M_2\equiv\frac{2\left(\lam_{2j}^2-\lam_{1j}^2\right)m_2}
{\left(2-\lam_{1j}^2\right)\left(2-\lam_{2i}^2\right)},
\label{m1m2} \ee where $M_1$ and $M_2$ are the effective masses of
the sources located at $z_1$ and $z_2$, respectively. A novel
feature of this configuration is that the function
$\omega_{\varphi}\equiv 2(A_1)_{\varphi}$ now contains information
about the spinning character of the gravitational field generated
by both sources since it has the following asymptotic behaviour
\be \omega_{\varphi}\mid_{\infty}\sim
\frac{2J_1\sin^2\theta_1}{r}+ \frac{2J_2\sin^2\theta_2}{r} \quad
\mbox{\rm with} \quad J_1=M_1\a_1, \quad J_2=M_2\a_2, \ee where
$J_1$ and $J_2$ define the angular momenta generated by the
spinning sources $M_1$ and $M_2$. This is a quite interesting
effect because of the dependence of $\omega_{\varphi}$ on both
rotational parameters $\a_1$ and $\a_2$. In fact, the
$G_{t\varphi}$--component of the seed field configuration did not
depend on $a_2$, however, the NHT induces a rotation in the charge
space which takes place in such a way that the source located at
$z_2$ gets as well an effective gravitational angular momentum
(apart from the rotating field described by the
$B_{t\varphi}$--component of the Kalb--Ramond tensor field).

On the other hand, we analyze the asymptotic behaviour of the
three--dimensional electric and magnetic potentials $A^I$ and
$s^T_I$ \be A^I\mid_{\infty}\sim\frac{Q_{2I}}{r}, \quad
s^T_I\mid_{\infty}\sim\frac{\mu_{1I}}{r}, \quad \mbox{\rm with}
\quad Q_{2I}\equiv
\frac{-4\left(m_1+m_2\right)\lam_{2I}}{\left(2-\lam_{2i}^2\right)},
\quad \mu_{1I}\equiv
\frac{4\left(m_2-m_1\right)\lam_{1I}}{\left(2-\lam_{1i}^2\right)},
\ee where $Q_{2I}$ and $\mu_{1I}$ determine the electric and
magnetic charges of the gauge fields, respectively.
\section{Conclusion and Discussion}

We have presented a double Ernst formulation of the stationary
bosonic string theory with some classes of exact solutions which
can be interpreted as a rotating field configuration generated by
two interacting sources of black hole type endowed with dilaton
and Kalb--Ramond fields. We clarify the rotating character of the
$B^{(4)}_{t\varphi}$--component of the antisymmetric tensor field
of Kalb--Ramond; this fact could be related to the {\it torsion}
nature of the antisymmetric background generated by the
Kalb--Ramond tensor field. Thus, string theory could predict the
necessary presence of torsion when studying gravitational field
configurations coupled or not to other matter fields. This line of
investigation deserves more attention and we hope to develop it in
the near future (for a review on this issue see, for example
\cite{torsion}).

By starting with a given field configuration of the stationary
effective field theory of the bosonic string, we obtained a new
configuration with the full field spectrum of the truncated
low--energy $D=4$ heterotic string theory via a straightforward
generation of the electromagnetic sector of the latter theory with
the aid of the non--linear matrix transformation of Harrison type.
We imposed as well the necessary condition on the charge vectors
in order to get asymptotically flat 4D gravitational field
configurations. Thus, in this way we have generated a field
configuration which consists of a pair of charged interacting
rotating black holes coupled to dilaton and Kalb--Ramond fields
and endowed with multiple electromagnetic charges (when $n=6$ the
field spectrum of the considered theory also describes the bosonic
sector of $D=\N=4$ supergravity).

In the framework of General Relativity the complete class of
solutions of the EM theory which describes an axisymmetric field
configuration consisting of a pair of black holes located on the
axis with arbitrary mass, charge and rotating parameters was
constructed in \cite{mmr} with an original method implemented by
Sibgatullin \cite{sibgatullin}. Due to its cumbersome form, it is
really difficult to extract particular solutions from it. We would
like to point out the main difference between the exact solution
presented here and that of \cite{mmr} (which gave rise to further
investigations \cite{et}--\cite{lt}). The point is that the
interaction between the black holes is different in both cases,
namely, in our configuration, the parameterization of the MEP $\X$
which leads to the double Ernst potential formulation of the
bosonic string theory fixes the linear superposition of the
sources (at the level of the Lagrangian), leading in turn to the
interaction given by
$e^{2\gamma}=e^{2(\gamma_{\epI}+\gamma_{\epII})}$ which is
different from the interaction considered in \cite{mmr} in the
framework of the Sibgatullin's method (see also \cite{cms}),
namely, the derivation of the solution involves the choice of a
plausible form of the Ernst potentials on the symmetry axis
($\rho=0,z$) and therefore extends these expressions to the whole
($\rho,z$) plane; the Ernst potential on the axis chosen in
\cite{mmr} reads \be \E(\rho=0,z)\equiv
e(z)=\frac{(z+z_1-m_1-ia_1)(z+z_2-m_2-ia_2)}
{(z+z_1+m_1-ia_1)(z+z_2+m_2-ia_2)} \ee whereas in the framework of
this method our choice corresponds to the potential \be
\epsilon_k(\rho=0,z)\equiv
e'(z)=\frac{(z-z_1-m_1+ia_1)}{(z-z_1+m_1+ia_1)}
+\frac{(z-z_2-m_2+ia_2)}{(z-z_2+m_2+ia_2)}; \ee this difference
leads to distinct functions $e^{2\gamma}$ and, thus, to different
interactions.

It should be pointed out that the integral equation method of Sibgatullin is more
powerful than the just algebraic method of nonlinear matrix transformations of
B\"acklund type. However, in the framework of string theory, one could choose more
general seed Ernst potentials than the (\ref{kerrpots}) ones and get therefore the
same interaction generated by the Sibgatullin method. This issue deserves more
attention and will be studied as well in a further work.

When certain parameters are set to zero, we recover some
previously studied configurations, however, one needs to perform a
great amount of algebraic computations in order to extract
explicit expressions for these solutions. We shall just point out
that when the Kalb--Ramond field and rotational parameters vanish
and we have just one electric charge at $z_1$ and $z_2$, we
reproduce a field configuration obtained by \cite{cp} and a
solution quite similar to the black dilohe considered in
\cite{emp}--\cite{et}.

We would like to emphasize as well that in the framework of General
Relativity the problem of equilibrium between two rotating black holes
has been intensively studied (see, for example, \cite{tk} and
references therein) and practically solved \cite{mrsg}--\cite{bmas}.

It is interesting to analyze other physical properties of the
constructed solution such as energy, equilibrium, thermodynamical
behaviour, dual configurations, etc. One can extract as well
gravitating monopole--antimonopole systems and uplift the
constructed solutions to 10 dimensions in order to reinterpret
them as brane--antibrane configurations. Some of these issues are
under current investigation.
\section{Acknowledgements}
The author acknowledges useful discussions with Deborah Aguilera
and Oleg Kechkin and thanks IMATE--UNAM and CINVESTAV for provided
library facilities while part of this investigation was in
progress. He is really grateful to S. Kousidou for encouraging him
during the performance of this article.

The author would like to express his gratitude to the Theoretical
Physics Department of the Aristotle University of Thessaloniki
and, specially, to Prof. J.E. Paschalis for fruitful discussions
and for providing a stimulating atmosphere while the final version
of this work was completed. He also acknowledges a grant for
postdoctoral studies provided by the Greek Government. This
research was supported by grants CIC-UMSNH-4.18 and
CONACYT-J34245-E and F42064.


\end{document}